\documentclass[aps,twocolumn,showpacs]{revtex4-1}
\usepackage{graphics,graphicx,amsfonts,amsmath,amsbsy,amssymb,color}
\usepackage[singlelinecheck=false]{subfig}
\usepackage{verbatim}  
\usepackage{psfrag}  
\usepackage{tabularx}
\usepackage{xmpmulti}
\usepackage{multirow}

\def \bfi {{\bf i}}
\def \bfj {{\bf j}}

\def \beq {\begin{eqnarray}}
\def \eeq {\end{eqnarray}}

\def \bfj {{\bf j}}
\def \bfi {{\bf i}}

\def \ket {{\rangle}}

\newcommand{\reffig}[1]{{Fig.~\ref{#1}}}

\newcommand{\half}{\frac{1}{2}}

\begin{document}

\title
{Emergence of Critical Phenomena in Full Configuration Interaction Quantum Monte Carlo}

\author{James~J.~Shepherd}
\email{js615@cam.ac.uk}
\affiliation{University of Cambridge, The University Chemical Laboratory, Lensfield Road, Cambridge, CB2 1EW, United Kingdom}
\author{Lauretta~R.~Schwarz}
\affiliation{University of Cambridge, The University Chemical Laboratory, Lensfield Road, Cambridge, CB2 1EW, United Kingdom}
\author{Robert~E.~Thomas}
\affiliation{University of Cambridge, The University Chemical Laboratory, Lensfield Road, Cambridge, CB2 1EW, United Kingdom}
\author{George.~H.~Booth}
\affiliation{University of Cambridge, The University Chemical Laboratory, Lensfield Road, Cambridge, CB2 1EW, United Kingdom}
\author{\\Daan Frenkel}
\affiliation{University of Cambridge, The University Chemical Laboratory, Lensfield Road, Cambridge, CB2 1EW, United Kingdom}
\author{Ali Alavi}
\email{asa10@cam.ac.uk}
\affiliation{University of Cambridge, The University Chemical Laboratory, Lensfield Road, Cambridge, CB2 1EW, United Kingdom}


\begin{abstract}

There has been recent literature discussion on the origin and severity of the `sign problem' in full configuration interaction quantum Monte Carlo (FCIQMC) and its `initiator' adaptation (\emph{i}-FCIQMC), methods of interest and potential because they allow for exact (FCI) ground-state solutions to be obtained often at a much reduced computational cost. In this study we aim to use a simple order parameter, describing the `sign structure' of the stochastic wavefunction representation, to empirically characterise the fundamentally different collective behaviour of the walker population in both methods.

\end{abstract}
\date{\today}
\maketitle

Full configuration interaction quantum Monte Carlo (FCIQMC)\cite{Booth2009}, and its initiator adaptation\cite{Cleland2010,Booth2011} (\emph{i}-FCIQMC), are electronic structure methods which attempt to iteratively solve the imaginary-time Schr\"odinger equation within a stochastic dynamic of walkers spanning a Slater determinant space. This approach has a great deal of potential because it has been shown to effectively diagonalize extremely large spaces and obtain ground-state energies, in molecules\cite{Booth2009,Booth2011,Cleland2010,Booth2010,Cleland2011,Umrigar2012unpub,Cleland2012unpub,Booth2012unpub,Daday2012} and a number of model systems\cite{Shepherd2012a,Shepherd2012b,Shepherd2012c,Spencer2012,Clark2012,Umrigar2012unpub}. 

Although this method is proving effective in a range of problems, analysis of why it is so successful at reducing the computational cost relative to FCI is still a topic of discussion\cite{Spencer2012,Clark2012a,Clark2012,Cleland2012unpub,Booth2012unpub,Kolodrubetz2012}. 
In this communication, we investigate the abrupt emergence of the `sign structure' of the wavefunction in FCIQMC to the exact solution by making an analogy with symmetry-breaking phase transitions and related critical point phenomena. This is contrasted with the smoother onset of order in the \emph{i}-FCIQMC method.
In the spirit of previous work analysing the so-called `sign problems' of other quantum Monte Carlo methods, it is hoped that work such as this to understand its origin and amelioration will lead to better strategies and lower cost in its treatment\cite{Foulkes2001,Cances2006,Luchow2000,Anderson1975,Anderson1976,Morales2012,Kalos1985,Coker1986,Anderson1991,Zhang1991,Kalos1996,Kalos2000,Kwon1998,Mitas2012}.

The crux of the FCIQMC method is that the finite-basis Schr\"odinger equation ($\hat{H}\Psi_0=E\Psi_0$; $\Psi_0 = \sum_\bfi C_\bfi | D_\bfi \ket$), is recast as a set of master equations governing the dynamics of a walker population in Slater determinant space,
\begin{equation}
-\frac{\text{d}C_\bfi}{\text{d}\tau} = (H_{\bfi\bfi}-E_{\text{HF}}-S)C_{\bfi} + \sum_{\bfj \neq \bfi} H_{\bfi\bfj}C_{\bfj},
\end{equation}
where the aim is to find $\frac{\text{d}C_\bfi}{\text{d}\tau}=0$ and recover $S$ as the correlation energy. The walkers themselves take a weight of $\pm1$\footnote{Although real formulations exist\cite{Umrigar2012unpub}} and in a converged calculation it is intended that they be distributed such that on average, $C_{\bf i} \propto N_{\bf i}$. These dynamics consist of different stochastically realized events, but in particular positively and negatively signed walkers are removed from the simulation by an `annihilation' criterion. 
Walkers of opposite signs simultaneously on the same site are removed at the end of each iteration, this procedure ensuring that each determinant has a definite sign once enough walkers have been introduced into the simulation.
This greatly ameliorates the FCIQMC `sign problem',
which has been investigated by Spencer, Blunt and Foulkes\cite{Spencer2012}, and is responsible for the required complexity of the instantaneous stochastic solution. They isolated a solution which the FCIQMC algorithm \emph{without sufficient annihilation} is unstable with respect to, characterised by the ground state of a modified Hamiltonian matrix, $H^\prime_{{\bf i} {\bf j}}=-|H_{{\bf i} {\bf j}}|$ for ${\bf i} \neq {\bf j}$. 

This explained the observation of a so-called annihilation plateau in FCIQMC: a `stalling' in population growth at a characteristic walker number ($N_c$). Above this walker number, the ground-state energy for the true Hamiltonian can typically be extracted, and below this walker number, the solution is contaminated by the aforementioned state. 

In the initiator adaptation to FCIQMC, an additional criterion is added to the step where walkers are created (spawning). The space is divided into those exceeding a certainly system-dependent parameter $n_\text{add}$, termed `initiators'. Spawning is conducted such that $H_{\bfi\bfj}$ is zeroed when spawning onto a site with no walkers, if the origin of the spawning attempt was not an `initiator'. The effect of this is to dynamically modify the Hamiltonian used to generate $\Psi\left(\tau + \delta\tau\right)$ from $\Psi\left(\tau\right)$. This can introduce an error at lower walker number due to the truncation of the instantaneously available space. Much of this error will be removed in the time-averaging, however systematic error can remain, and is referred to as the `initiator error'.

An FCIQMC or \emph{i}-FCIQMC simulation in general begins with a single, positively-signed walker on a reference determinant (typically the HF determinant)\footnote{Or made into a multi-determinantal expansion\cite{Umrigar2012unpub}}. From this point, the population is grown by setting the shift parameter to be fixed such that $S>E_0$. Typically, at some point in the simulation the population is relaxed from this fixed-shift mode, and the shift is updated in a self-consistent fashion\footnote{Although the relationship between these two modes is discussed elsewhere\cite{Shepherd2012b}}.

\begin{figure*}
\subfloat[]{\includegraphics[width=0.32\textwidth]{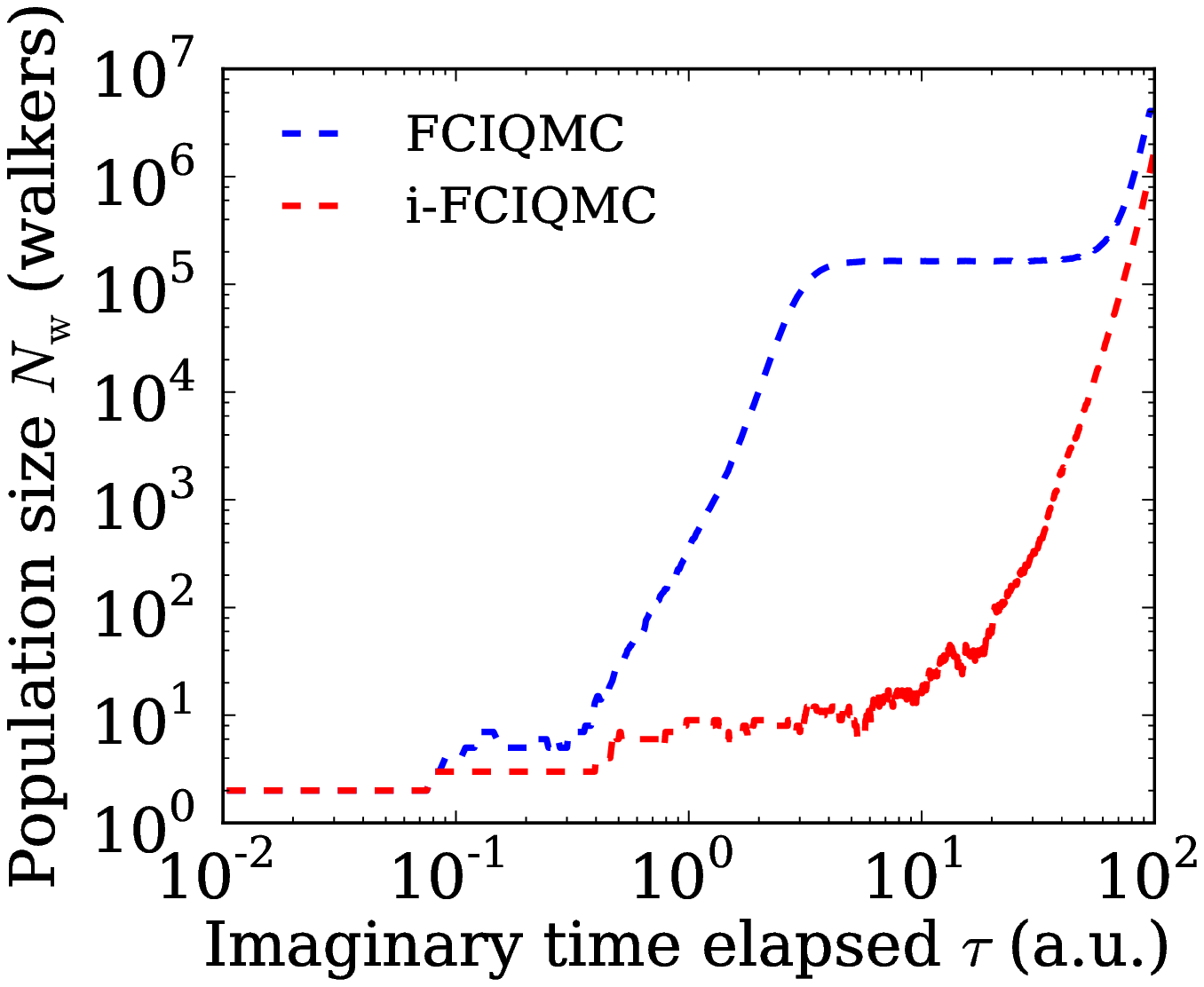}\label{plateau}}
\subfloat[]{\includegraphics[width=0.32\textwidth]{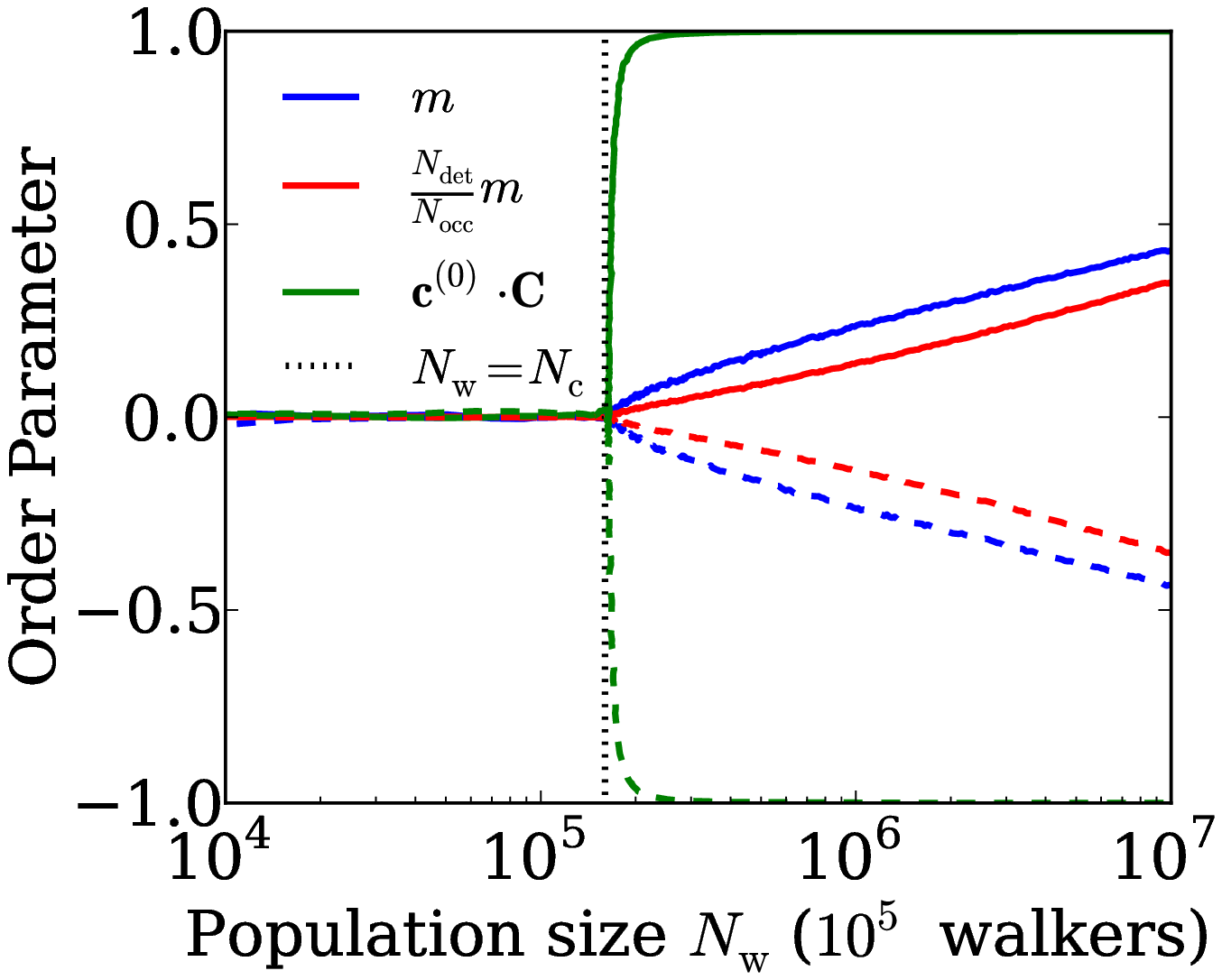}\label{order_param}}
\subfloat[]{\includegraphics[width=0.32\textwidth]{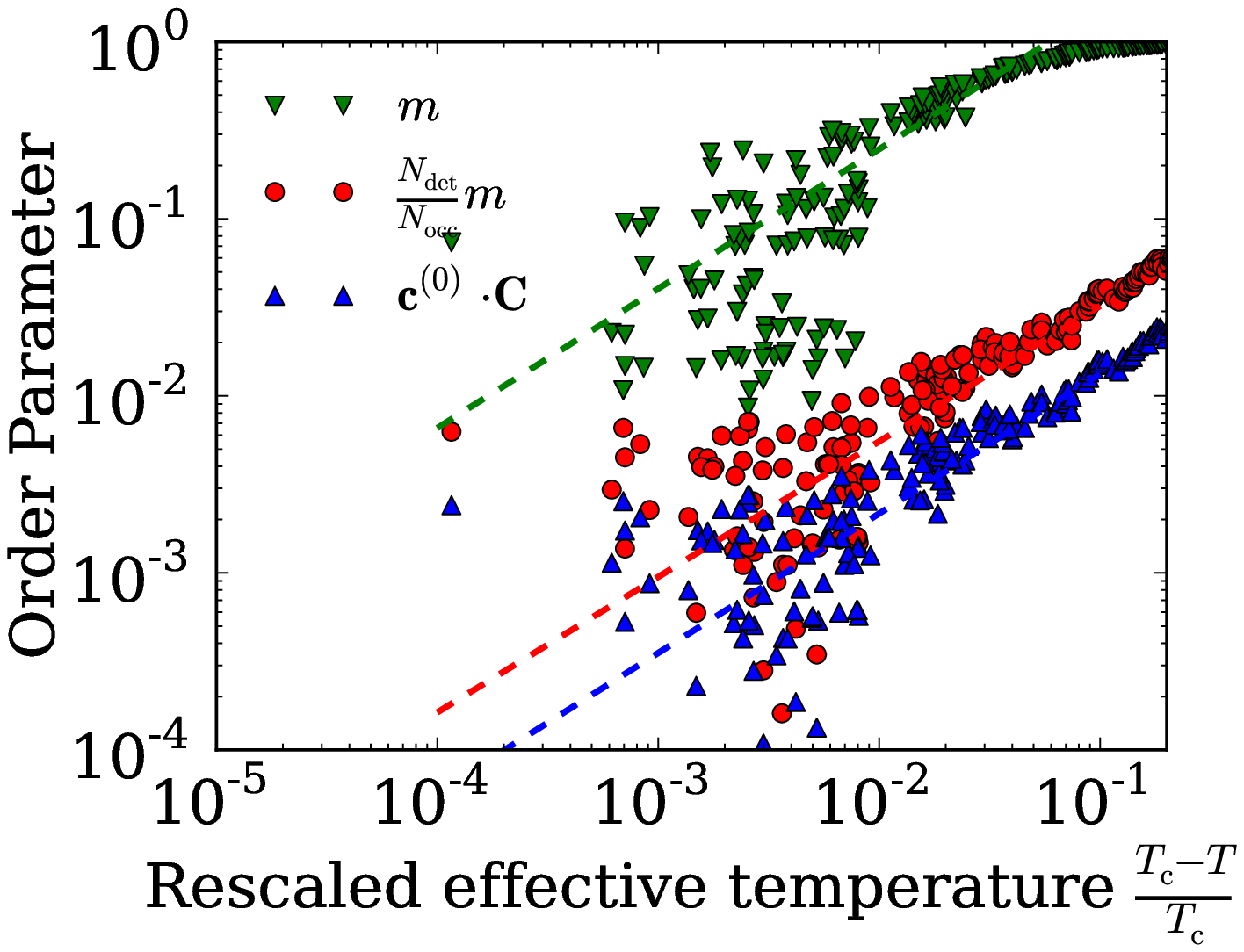}\label{phase_trans}}
\caption{(a) In FCIQMC, there naturally arises a plateau in walker growth termed the annihilation plateau ($N_\text{c}=N_\text{w}\sim164,000$). This feature is absent in \emph{i}-FCIQMC. (b) Above this critical walker number, there is a change in the order parameter from oscillating about zero to having a non-zero value. The sign is trajectory-dependent (dashed lines correspond to the same parameters as the solid lines, with a different trajectory). The speed at which this reaches $\pm1$ is dependent on the order parameter (discussed in the text). (c) Transforming $1/N_\text{w}$ to an effective temperature ($T$), the critical exponent for small values of the order parameter are found to be $\beta=0.77\pm0.03$.}
\label{}
\end{figure*}

We chose as our model system the beryllium dimer at a near-equilibrium geometry of 3.8329 a.u., in a cc-pVTZ basis set, with frozen core electrons. 
Our molecular integrals were calculated by the {\tt QCHEM} package\cite{Qchemshort} ($E_\text{HF}=-29.1123$~a.u.). This was chosen predominantly because it is a system with a small determinant space of $N_\text{FCI}=346\, 485$, with a comparatively large plateau height of $N_\text{c} \sim 164\, 000$ walkers (there is a narrow gaussian distribution of $N_\text{c}$ between different simulations, $S$=0.0~a.u., with a time step of 0.001~a.u.). 
It is also an example of a system trivially solved by \emph{i}-FCIQMC with very little initiator error at modest walker numbers. A typical simulation profile for FCIQMC, showing the growth of the walker population, is shown in \reffig{plateau}. At approximately $\tau=3$, the population enters the so-called annihilation plateau and stabilises for a certain length of time before exiting (around $\tau=50$). The length of time spent in the plateau varies between simulations and has an exponential distribution in its tail. No such plateau exists in an equivalent initiator simulation for Be$_2$ and for all molecular systems studied to date. 

We seek a single parameter which we can use to monitor the convergence of the instantaneous stochastic wavefunction representation,
\begin{equation}
\Psi\left(\tau\right)=\sum_{\bf i} C_{\bf i} | D_{\bf i} \rangle
\end{equation}
to the true ground-state (FCI) wavefunction
\begin{equation}
\Psi_0=\sum_{\bf i} c_{\bf i}^{(0)} | D_{\bf i} \rangle.
\end{equation}
or the degenerate alternative $-\Psi_0$, where ${\bf c}^{(0)}$ and ${\bf C}$ are normalised. Motivated by discussion of `sign structure' in previous papers we construct the following parameter,
\begin{equation}
m=\frac{1}{N_\text{det}} \sum_{\bf i}^\prime \frac{C_{\bf i}}{|C_{\bf i}|}  \frac{c_{\bf i}^{(0)}}{|c_{\bf i}^{(0)}|},
\end{equation}
where $N_\text{det}$ is the number of non-zero weighted determinants in $\Psi_0$, and the prime above the sum indicates that the sum should not be taken over any component where the numerator is zero. This expresses the proportion of the signs that are correctly-aligned with the final wavefunction, with the following interpretation:
\begin{equation}
m \left\{ \begin{array}{ll}
= 1,  & \text{signs completely aligned with } +\Psi_0 \\
> 0, & \text{signs more aligned with } +\Psi_0 \text{ than } -\Psi_0 \\
= 0, & \text{randomly aligned signs} \\
< 0, & \text{signs more aligned with } -\Psi_0 \text{ than } \Psi_0 \\
= -1, & \text{signs completely aligned with } -\Psi_0.
\end{array}
\right.
\end{equation}
Due to the similarity with order parameters in many branches of statistical physics we can also consider this as such.
We will also employ an \emph{occupied-space} order parameter in this study, where the parameter is re-weighted by $N_\text{det}/N_\text{occ}$, where $N_\text{occ}$ is the number of non-zero weighted determinants in $\Psi\left(\tau \right)$. We will simply denote this $\frac{N_\text{det}}{N_\text{occ}} m$. This is in general larger than $m$ (although has the same limiting values of $\pm1$) and has the effect of only finding the ratio of sign alignment with in the space which has walkers in it, rather than over the whole space, and in particular has a starting-point of $\pm1$ rather than $\pm \frac{1}{N_\text{det}}$. We will also compare these with an overlap value ${\bf c^{(0)}}\cdot{\bf C}$, which includes consideration of the magnitude as well as sign overlap of the wavefunctions.

For FCIQMC simulations on the beryllium dimer directly corresponding to \reffig{plateau}, calculated order parameters are shown in \reffig{order_param} with respect to different walker populations, $N_\text{w}$. We take the approach of only showing the instantaneous order parameter of a single trajectory at a time, rather than attempting to deal with stochastic error, noting that these are extremely similar to data from averages taken by going into so-called variable shift mode. This is because the equilibration time for this system is close to zero and the qualitative behaviour at each point insensitive to the starting random-number sequence. This situation is discussed extensively in another paper by the authors\cite{Shepherd2012b}.

The rising magnitude of the order parameter results from gradual rise in the overlap between the simulation wavefunction and exact wavefunction (${\bf c^{(0)}}\cdot{\bf C}$, also shown in \reffig{order_param}). In particular, the graph shows two regions. 

For $N_\text{w} < N_\text{c}$, the order parameters fluctuate about zero. This represents no average net alignment of the signs, or overlap, with either of $+\Psi_0$ or $-\Psi_0$. In pre-plateau simulations, or simulations without sufficient annihilation, the ground-state solution is contaminated by a solution from an effective Hamiltonian ($H^\prime_{{\bf i} {\bf j}}=-|H_{{\bf i} {\bf j}}|$ as described earlier\cite{Spencer2012}). Since the effective Hamiltonian only has negative off-diagonal matrix elements, the resultant eigenvector has same signed coefficients, although there is still a degenerate solution corresponding to a global sign change. The bulk of the positively (negatively) signed walkers are distributed according to the positively (negatively) signed solution in this phase, resulting in no net instantaneous overlap with $\pm\Psi_0$. This is the region in which the sign of the reference determinant can be seen to flip in some systems\cite{Clark2012}.

For $N_\text{w} > N_\text{c}$, the order parameters become non-zero, showing significant emergence of either $-\Psi_0$ or $+\Psi_0$ as the dominant solution (we have shown an example of both types of trajectory). 
The growth in this parameter $m$ is then apparently logarithmic in walker number. This contrasts the true overlap 
which grows extremely rapidly. The rapid growth in this overlap causes the projected energy and shift estimators for the correlation energy to be exact beyond exit from the plateau\cite{Booth2009,Booth2010,Cleland2010,Booth2011}. 

The discontinuity at $N_\text{w}=N_\text{c}$ corresponds directly to \emph{entry} into the plateau region.
In fact, the overlap order parameter ${\bf c^{(0)}}\cdot{\bf C}$ grows to $\sim$0.6 before the a clear exit from the plateau is seen, a value which fluctuates strongly with $N_\text{w}$ and imaginary time.

The slow approach of the order parameter $m$ to unity is due to the instantaneous wavefunction bearing little resemblance to the final wavefunction $\Psi_0$ for low-amplitude determinants, only being \emph{on average} able to yield these coefficients\cite{Cleland2012unpub}. Stochastic fluctuations therefore not only modify the coefficients of the eigenvector but also the sign structure. In principle, we could average the wavefunction over imaginary time, whereupon we would expect the average to approach $\Psi_0$. 

Using our physical order parameter analogy, we can think of the walker population $N_\text{w}$ as being related to an effective temperature scale $T=1/\left( N_\text{w} \right)$. The plateau height $N_\text{c}$ can be seen to be a critical temperature $T_{c}$, and at this critical temperature the population settles into an ordered phase of either $+\Psi_0$ or $-\Psi_0$. Ergodicity is broken and the other ordered phase is not explored beyond entry into the plateau. This phase transition, characterised by our order parameter, can be seen to be second-order with a critical exponent defined by,
\begin{equation}
m \sim \left( \frac{ T_\text{c} - T }{ T_\text{c} } \right)^\beta,
\end{equation}
with $\beta\sim=0.77\pm0.03$ (\reffig{phase_trans}). The region of fitting is the linear region taken by the distribution of points beyond $10^{-2}$, and varying the plateau height by $\pm2\,000$ (from $N_\text{c}=164\,000$) causes deviation in our estimate of $\beta$ by at most $\pm0.05$. This $\beta$ is consistent with the data from overlap order parameter ${\bf c^{(0)}}\cdot{\bf C}$, and although cannot be independently found from a free-fit due to noise in the data demonstrates that measurement of $\beta$ is consistent between different measures of this ordering.

There remains much to be understood about the meaning of this $\beta$ parameter and a complete investigation of this is well beyond the scope of this initial study. We speculate that its value being \emph{greater} than a (Landau) mean-field value of $\beta=\half$ means that there is anti-cooperativity between the presently ordered phase and the surrounding space; a new determinant needs to be consistently ordered with respect to a growing set of ordered determinants. We expect it would be possible to link this to behaviour in nature, and note that this value is similar to that observed by simulations of frustrated spin-glasses\cite{Young1988}. Although universal between different methods of measuring the ordering, we would expect it to be strongly system-dependent and hence might be a measure of how intrinsically entangled the wavefunction is.

\begin{figure*}
\subfloat[]{\includegraphics[width=0.32\textwidth]{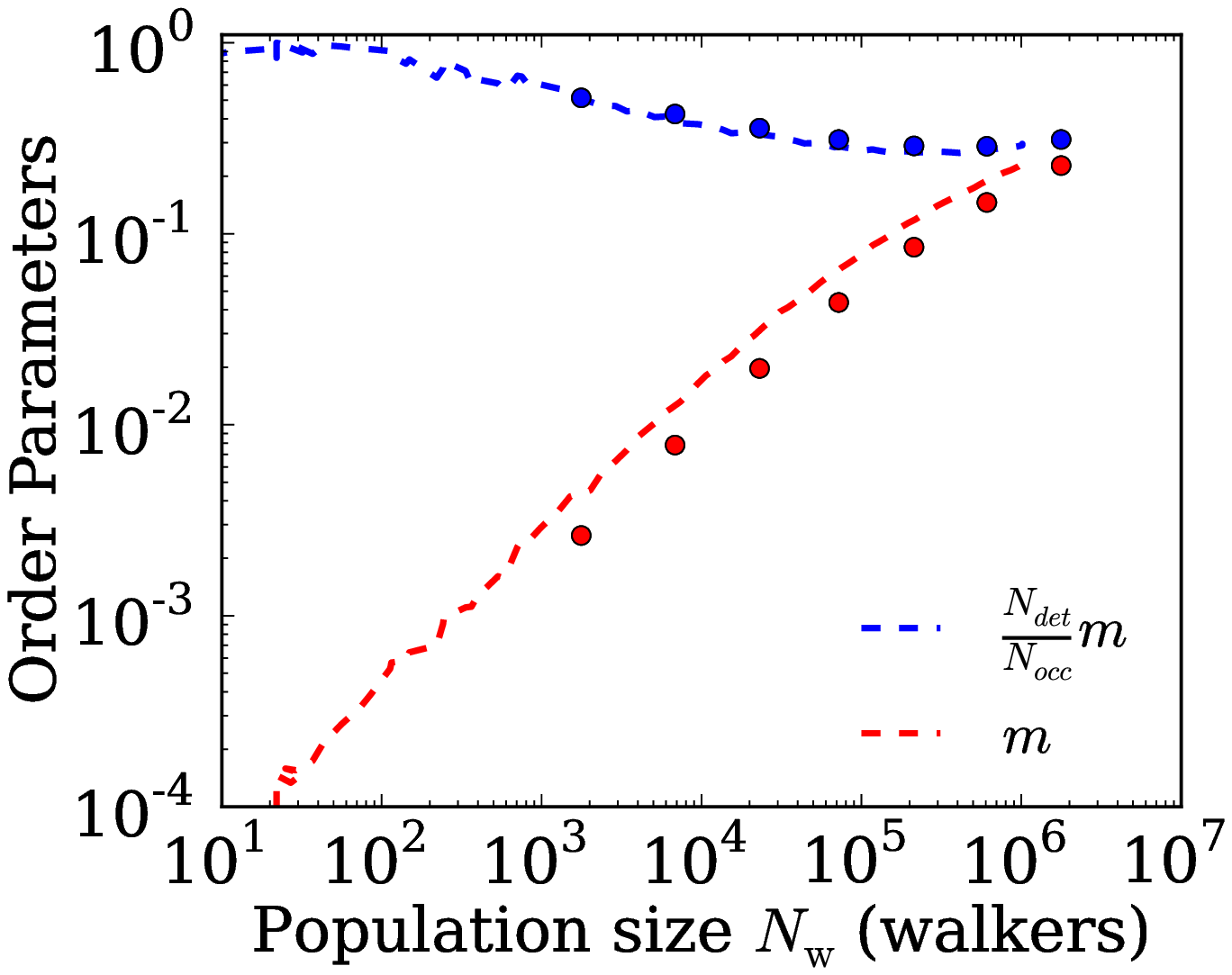}\label{init_order_param_inst}}
\subfloat[]{\includegraphics[width=0.32\textwidth]{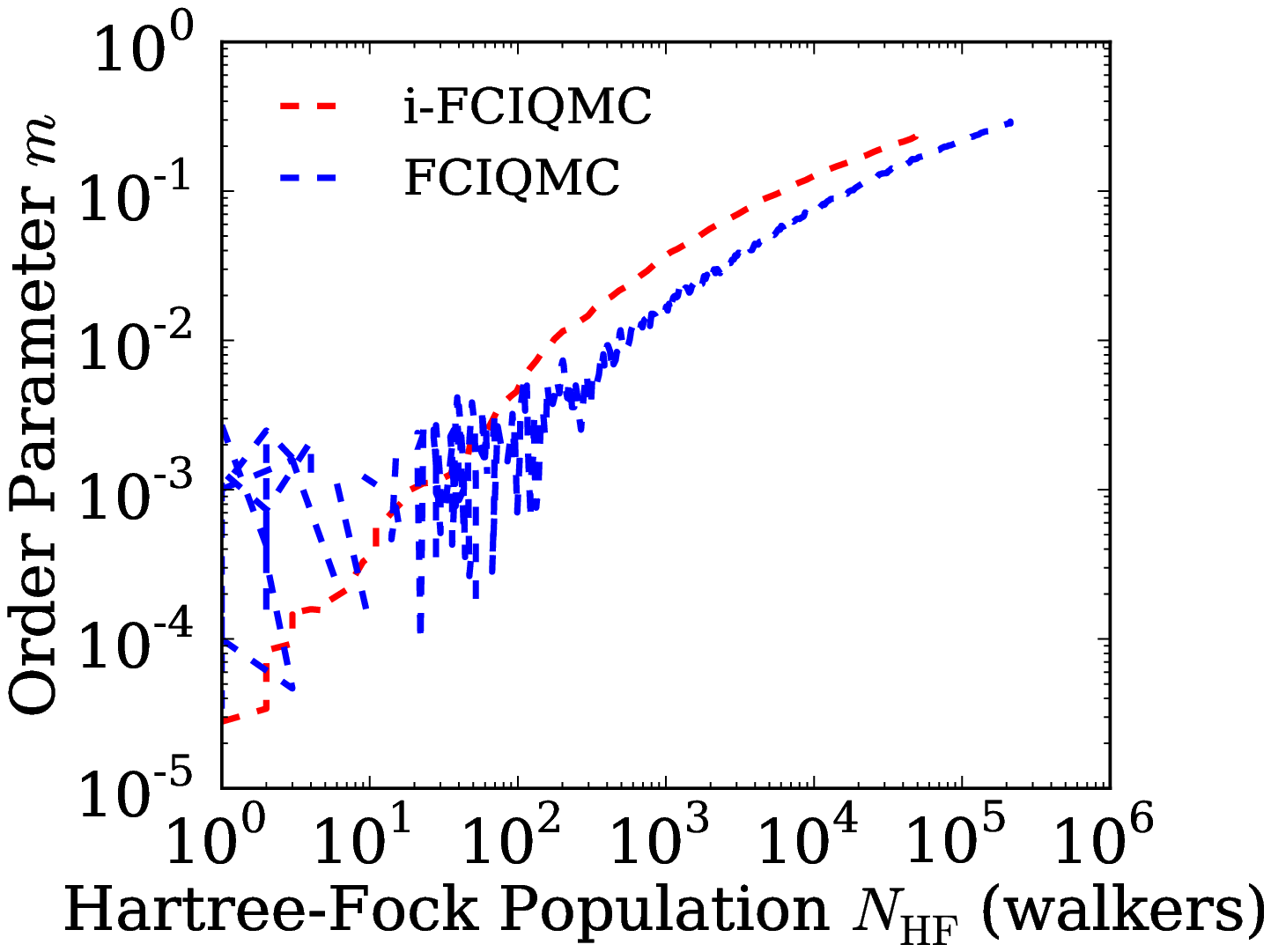}\label{similarity}}
\subfloat[]{\includegraphics[width=0.32\textwidth]{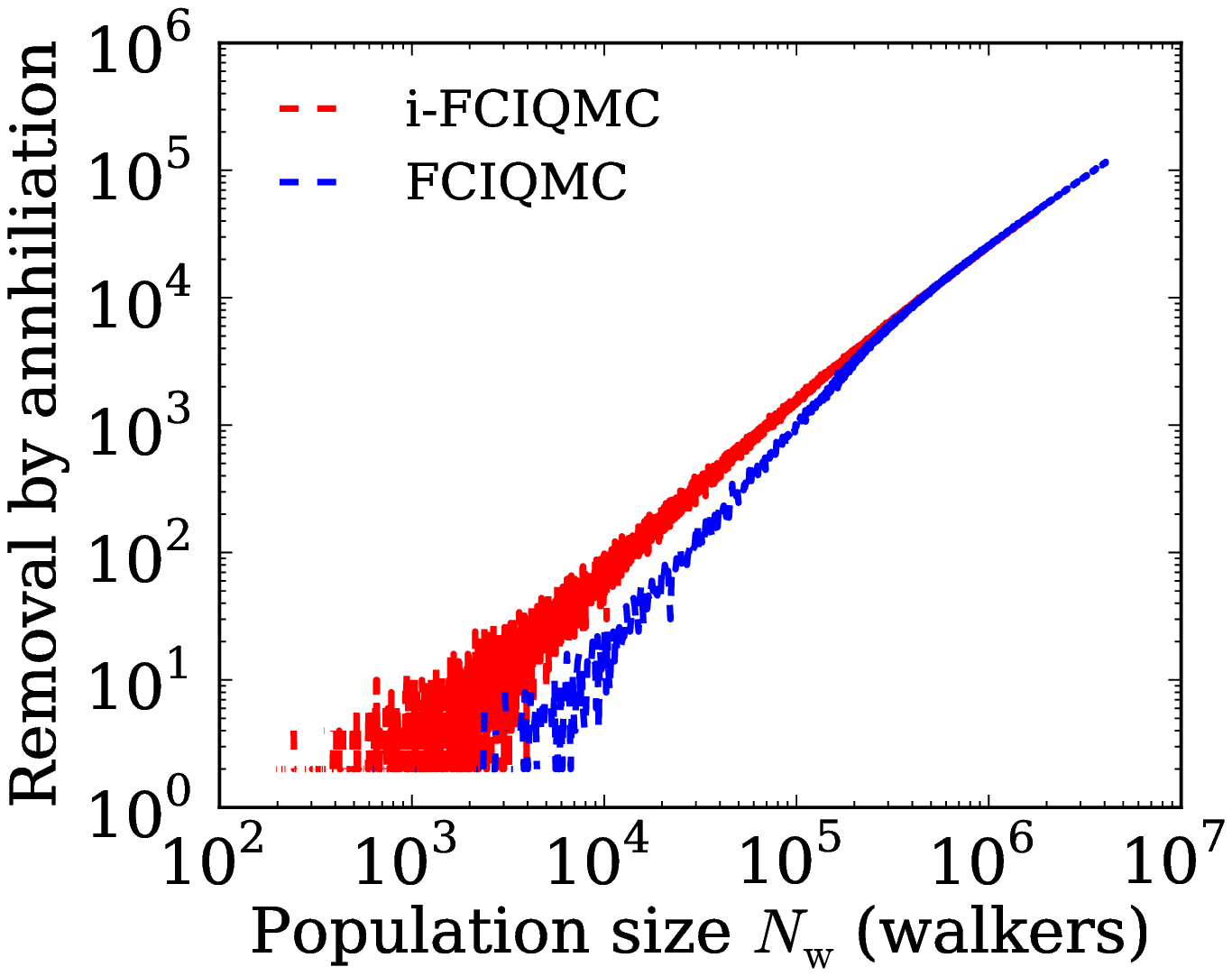}\label{annhil}}
\caption{(a) In \emph{i}-FCIQMC, there is no annihilation plateau and the order parameter does not change sign, remaining at the same sign as the original walker distribution (positive in this case). There is immediate monotonic growth of the whole-space order parameter $m$. In contrast, the reduced-space order parameter initially falls, due to increasing distribution of the walker population across the space, before rising as this is overtaken by these signs aligning. Filled circles represent averaged values taken by going into variable shift mode, whereas the dashed line traces out the instantaneous values during walker growth. (b) When the order parameter is plotted with respect to the population on the Hartree--Fock determinant, FCIQMC and \emph{i}-FCIQMC are much more similar. (c) For a given population size, there is greater annihilation in \emph{i}-FCIQMC than FCIQMC.}
\label{}
\end{figure*}

Having shown that our order parameter $m$ can produce the anticipated behaviour for FCIQMC, we turn our attention to the initiator adaptation of FCIQMC (\emph{i}-FCIQMC). We begin by noting that Be$_2$ is almost immediately solved for almost any walker number in \emph{i}-FCIQMC. The dimer is actually an extreme example of a difficult system for FCIQMC, and a straight-forward system for \emph{i}-FCIQMC. 

In \reffig{init_order_param_inst}, the same order parameter analysis is repeated for the initiator approximation. Here, averaged values obtained in variable shift mode are shown by solid circle on the plot, and the instantaneous and averaged values overlay one another with a slight shift in $N_\text{w}$. This mismatch is anticipated, and is due to $S\neq E_{\text{corr}}$ as described in Ref. \onlinecite{Shepherd2012b}. 

The order parameters for \emph{i}-FCIQMC are very different to those for FCIQMC. The whole-space order parameter $m$, rises continuously remaining positive throughout. Correspondingly, there is no sign-flipping of the reference determinant as in pre-plateau FCIQMC simulations for this system. The occupied-space order parameter, $\frac{N_\text{det}}{N_\text{occ}} m$, also remains positive for the same reason. However, it falls from its initial starting point of $+1$ (or in general $\pm1$), as the walkers distribute themselves across the space at a quicker rate than sign coherence can spread. The latter then catches up as more of the space becomes aligned. The two order parameters $m$ and $\frac{N_\text{det}}{N_\text{occ}} m$ differ significantly in \emph{i}-FCIQMC whilst they do not differ by much in FCIQMC. This is in part because the occupied space in \emph{i}-FCIQMC is typically (always for Be$_2$) smaller than the occupied space for FCIQMC. 

Finally, \reffig{similarity} shows that, from the point of view of the population at the Hartree--Fock determinant, there is a remarkable similarity between order parameters in FCIQMC and \emph{i}-FCIQMC. This has been observed previously as the idea that the HF determinant only grows its population in the post-plateau phase of the calculation\cite{Thom2010}, and further relates the symmetry-broken phase of FCIQMC to the whole of an \emph{i}-FCIQMC simulation.

These data are consistent with the idea that the annihilation plateau is removed by the `initiator' adaptation for systems in which it is effective, which can also been seen in enhanced annihilation rates in \emph{i}-FCIQMC compared with FCIQMC (\reffig{annhil}). We agree with Spencer \emph{et al.}\cite{Spencer2012} that the restriction of the simulation to a sub-space imposed by linking $\Psi \left( \tau \right)$ and $H_{{\bf i}{\bf j}} \left( \tau \right)$ leads to increased annihilation because the walkers are able to `meet' each other more. On the other hand, there is no non-trivial link between this and a sub-space diagonalization, since energies are much-improved compared with those that are produced by a diagonalization in the occupied subspace\cite{Cleland2012unpub}. Looking at the beginning of the simulation, there is much slower growth in \emph{i}-FCIQMC than in FCIQMC (\reffig{}), which is because walkers are being admitted at a much slower rate into the simulation. This restriction, and the smooth growth in the order parameter, happens at a much lower walker number than significant annihilation begins. 

In summary, we have introduced a parameter ($m$) which measures the extent of agreement between the signs of the stochastic wavefunction representations in FCIQMC and \emph{i}-FCIQMC and the true ground-state solution. When appropriately normalised, this measures $\pm1$ for perfect sign agreement with $\pm\Psi$. In FCIQMC simulations of the beryllium dimer, $m$ fluctuates about zero for small populations, taking both positive and negative values as the sign of the reference determinant (and all determinants) flips. After the annihilation plateau is entered, for a single trajectory the sign of $m$ is `locked in' to either approach $+1$ or $-1$ as the walker population is grown. This is linked to a phase transition at an effective temperature given by an inverse walker number. In contrast, \emph{i}-FCIQMC has no annihilation plateau, sign flipping, or observed phase transition behaviour in the order parameters, instead immediately growing towards $+1$. We relate to increased annihilation in the system due to the dynamical Hamiltonian, which contrains the instantaneous space and enhances the removal of the sign problem in \emph{i}-FCIQMC by annihilation. 

As with any empirical or numerical study of such phenomena, one important issue to address is the transferability of these findings. In particular, Be$_2$ is a system with a pronounced plateau in FCIQMC, but that is largely problem-free in \emph{i}-FCIQMC. We have effectively, therefore, used it to show how we believe \emph{i}-FCIQMC to be working when it works at its best. In particular, these conclusions would probably not be true in an \emph{i}-FCIQMC calculation that exhibited an annihilation plateau or severe sign-flipping and such simulations lie beyond the scope of this paper.

The authors thank Mark A. Miller for valuable discussions and gratefully acknowledge Trinity College (GHB) and EPSRC (JJS, RET, AA) for funding.

\bibliography{bib}

\end{document}